# Equivalence Principles, Lense-Thirring Effects, and Solar-System Tests of Cosmological Models

Wei-Tou Ni[1,2]

[1]Center for Gravitation and Cosmology (CGC), Department of Physics,
National Tsing Hua University, Hsinchu, Taiwan, 30013 Republic of China

[2]Shanghai United Center for Astrophysics (SUCA),
Shanghai Normal University, Shanghai, 20234 China

*Email*: weitou@gmail.com

**Abstract**

In this talk, we review the empirical status for modern gravitational theories with emphases on (i) Equivalence Principles; (ii) Lense-Thirring effects and the implications of Gravity Probe B experiment; (iii) Solar-System Tests of Cosmological Models.

## 1 Empirical Foundations of the Einstein Equivalence Principle

Einstein Equivalence Principle (EEP) is a cornerstone of general relativity and metric theories of gravity. For examining its empirical foundation, we need a framework embracing general relativity. For doing this, we use the $\chi$-$g$ framework (Ni, 1983, 2010) which is summarized in the following interaction Lagrangian density

$$L_I = - (1/(16\pi))\chi^{ijkl} F_{ij} F_{kl} - A_k j^k (-g)^{(1/2)} - \Sigma_I m_I (ds_I)/(dt)\, \delta(\boldsymbol{x}\text{-}\boldsymbol{x}_I), \tag{1}$$

with $\chi^{ijkl} = \chi^{klij} = -\chi^{jikl}$ a tensor density of the gravitational fields (e.g., $g_{ij}$, $\varphi$, etc.) or fields to be investigated. The gravitational constitutive tensor density $\chi^{ijkl}$ dictates the behaviour of electromagnetism in a gravitational field and has 21 independent components in general. For general relativity or a metric theory (when EEP holds), $\chi^{ijkl}$ is determined completely by the metric $g_{ij}$ and equals $(-g)^{1/2}[(1/2)g^{ik}g^{jl}-(1/2)g^{il}g^{jk}]$; when $g^{ik}$ is replaced by $\eta^{ik}$, we obtain the special relativistic Lagrangian density. The SME (Standard Model Extension; Kostelecky and Mews, 2002) and SMS (Standard Model Supplement; Ma 2012) overlap the $\chi$-$g$ framework in their photon sector. Hence, our studies are directly relevant to parameter constraints in these models.

In the following, we summarize experimental constraints on the 21 degrees of freedom of $\chi^{ijkl}$ to see how close we can reach EEP and metric theory empirically.

*Constraints from no birefringence*: In the χ-g framework, the theoretical condition for no birefringence (no splitting, no retardation) for electromagnetic wave propagation in all directions is that the constitutive tensor $\chi^{ijkl}$ can be written in the following form

$$\chi^{ijkl} = (-H)^{1/2}[(1/2)H^{ik}H^{jl} - (1/2)H^{il}H^{kj}]\psi + \varphi e^{ijkl}, \qquad (2)$$

where $H = \det(H_{ij})$ and $H_{ij}$ is a metric which generates the light cone for electromagnetic propagation, and $e^{ijkl}$ the completely antisymmetric symbol (Ni, 1984; Lämmerzahl and Hehl 2004). Polarization measurements of light from pulsars and cosmologically distant astrophysical sources yield stringent constraints agreeing with (2) down to $2 \times 10^{-32}$ fractionally; for a review, see Ni (2010).

With (2), we still have a possibly new photon metric $H_{ij}$, an axion degree of freedom, $\varphi e^{ijkl}$, and a 'dilaton' degree of freedom, $\psi$. To fully recover EEP, we need (i) good constraints on only one physical metric, (ii) good constraints on no $\psi$ ('dilaton'), and (iii) good constraints on no $\varphi$ (axion) or no pseudoscalar-photon interaction. *Good constraints on one physical metric and no 'dilaton' ($\psi$)* come from Hughes-Drever-type experiments, Eötvös-Dicke-type experiments and redshift experiments. For a detailed account, please see Ni (2010).

With constraints from (i) no birefringence, (ii) no extra physical metric, (iii) no $\psi$ ('dilaton'), we arrive at the theory (1) with $\chi^{ijkl}$ given by

$$\chi^{ijkl} = (-g)^{1/2}[(1/2)g^{ik}g^{jl} - (1/2)g^{il}g^{kj} + \varphi \varepsilon^{ijkl}], \qquad (3)$$

i.e., an axion theory (Ni, 1984; Hehl and Obukhov 2008). Here $\varepsilon^{ijkl} \equiv (-g)^{-1/2} e^{ijkl}$. The current constraints on the cosmic polarization rotation angles $\Delta\varphi$'s are within $\pm$ 30 mrad from astrophysical observations and CMB polarization observations (Ni, 2010). Thus, from experiments and observations, only one degree of freedom of $\chi^{ijkl}$, i.e., the $\varphi$ degree of freedom is not much constrained.

Now let's turn into more formal aspects of equivalence principles. We proved that for a system whose Lagrangian density given by equation (1), the Galileo Equivalence Principle (UFF [Universality of Free Fall; Weak Equivalence Principle I (WEP I)]) holds if and only if equation (3) holds (Ni, 1974, 1977).

If $\varphi \neq 0$ in (3), the gravitational coupling to electromagnetism is not minimal and EEP is violated. Hence WEP I does not imply EEP and Schiff's conjecture (which states that WEP I implies EEP) is incorrect (Ni, 1973, 1974, 1977). However, WEP I does constrain the 21 degrees of freedom of χ to only one degree of freedom (φ), and Schiff's conjecture is largely right in spirit. To fully imply EEP, we also need WEP II, which also assumes no anomalous torques (Ni, 1974, 1977).

The theory with $\varphi \neq 0$ is a pseudoscalar theory with important astrophysical and cosmological consequences. This is an example that investigations in fundamental physical laws lead to implications in cosmology (Ni, 1977). Investigations of CP problems in high energy physics lead to a theory with a similar piece of Lagrangian with $\varphi$ the axion field for QCD (Peccei and Quinn, 1977; Weinberg, 1978; Wilczek, 1978).

In this section, we have shown that EEP and the empirical foundations of classical electromagnetism are solid except in the aspect of a possible pseudoscalar photon interaction. From here, we have built a Parametrized Post-Maxwellian (PPM) framework for testing electromagnetism with this piece and quantum corrections.

## 2  Lense-Thirring Effects, GP-B and WEP II

Lense and Thirring (1918) showed the dragging of the orbit plane of a satellite around a rotating planet in general relativity. Schiff (1960) showed that an ideal gyroscope in orbit around Earth would undergo two relativistic precessions with respect to a distant inertial frame: (i) a geodetic drift in the orbit plane due to motion through the space-time curved by Earth's mass and (ii) a frame-dragging due to Earth's rotation. The geodetic term matches the curvature precession of the Earth-Moon system around the Sun given by de Sitter (1916). The Schiff frame dragging is related to Lense-Thirring dragging of the orbit plane. Lense-Thirring effects/frame dragging have important implications for astrophysics; it has important effects in the formation of accretion disk (Bardeen and Petterson, 1975) and has been invoked as a mechanism to drive relativistic jets emanating from galactic nuclei (Thorne, 1988).

GP-B, a space experiment launched 20 April 2004, with 31 years of research and development, 10 years of flight preparation, a 1.5 year flight mission and 5 years of data analysis, had arrived in 2011 at the final experimental results for this landmark testing two fundamental predictions of Einstein's theory of General Relativity (GR), the geodetic and frame-dragging effects, by means of cryogenic gyroscopes in Earth orbit. The spacecraft carries 4 gyroscopes (quartz balls) pointing to the guide star IM Pegasi in a polar orbit of height 642 km. GP-B was conceived as a controlled physics experiment having mas/yr stability ($10^6$ times better than the best modeled navigation gyroscopes) with numerous built-in checks and methods of treating systematics. Analysis of the data from all four gyroscopes results in a geodetic drift rate of −6601.8 ± 18.3 mas/yr and a frame-dragging drift rate of −37.2 ± 7.2 mas/yr, to be compared with the GR predictions of −6606.1 mas/yr and −39.2 mas/yr. The Schiff frame dragging is verified with an uncertainty of 19 %. The de Sitter geodetic precession is verified to 0.28 %. (Everitt et al., 2011)

With precise Satellite Laser Ranging (SLR) to LAGEOS satellite and with GRACE geodesy data, Lense-Thirring prediction on orbit precession in general relativity has been verified to 10-30 % (Ciufolini and Pavlis, 2004; Iorio, 2009). It is gratifying that the LAGEOS measurement and the GP-B measurement of the two related Lense-Thirring effects are similar in accuracy and they agree with each other and with general relativity.

In the GP-B experiment, the 4 quartz balls are in drag-free free fall and can be used to test WEP II with bodies in different rotation state. The free-fall trajectories are equivalent and independent of rotation state to very high precision with the free-fall Eötvös parameter $|\eta| \leq 10^{-11}$ which is a four-order improvement over previous results for rotating body (Ni, 2011). We use GP-B results also to test WEP II for whether there is an anomalous torque; the anomalous torque per unit angular momentum parameter λ is constrained to $(-0.05 \pm 3.67) \times 10^{-15}$ s$^{-1}$, $(0.24 \pm 0.98) \times 10^{-15}$ s$^{-1}$, and $(0 \pm 3.6) \times 10^{-13}$ s$^{-1}$ respectively in the directions of geodetic effect, frame-dragging effect and angular momentum axis; the dimensionless frequency-dependence parameter κ is constrained to $(1.75 \pm 4.96) \times 10^{-17}$, $(1.80 \pm 1.34) \times 10^{-17}$, and $(0 \pm 3) \times 10^{-14}$ respectively (Ni, 2011).

# 3 Solar-System Tests of Cosmological Models

In this section, we use the solar-system measurements of the relativistic time delay and light deflection to constrain the quadratic Weyl term parameter (Ni, 2012) in the DSSY theory (Derulle *et al*., 2011), and use the prospects of ranging measurement of relative perihelion shifts to project on the possible future measurements/constraints (Ni, 2009) on the DGP theory (Dvali *et al*., 2000) and massive gravity (de Rham and Gabadadze, 2010). First, we summarize the relevant solar-system tests.

## 3.1 Solar-system tests

The solar-system tests are summarized in Ni (2005) up to 2005. Recent measurements of PPN parameter γ and the ranging accuracy are discussed in the following.

### 3.1.1 Measurement of PPN parameter γ

PPN parameter γ is measured in Shapiro time delay experiments, radio wave deflection experiments and GP-B experiment. Table 1 compiles the most recent results of these experiments.

Table 1. Recent solar-system measurements of the PPN space curvature parameter γ.

| Quantity measured | Experiment | Value and Uncertainty |
|---|---|---|
| Shapiro time delay | Cassini S/C Ranging (Bertotti *et al*., 2003) | 1.000021±0.000023 |
| Radio wave deflection | VLBI solar deflection (Shapiro *et al*., 2004)<br>VLBA solar deflection (Fomalont *et al*., 2009)<br>Geodetic VLBI since 1979 (Lambert *et al*., 2009) | 0.99983±0.00045<br>0.9998±0.0003<br>1.0000±0.0002 |
| Geodetic Precession | Gravity Probe B (Everitt *et al*., 2011)<br>(Earth de Sitter effect) | 0.99935±0.0028 |
| | Lunar Laser Ranging (Williams *et al*., 2004)<br>(Solar de Sitter effect) | 0.9981±0.0064 |

The error of these experiments is quoted in terms of PPN γ in Table 1. Since the effects are proportional to $(1 + \gamma)$, the agreements with general relativistic effects are within half of the value of the errors for γ. The most precise experiment of relativistic time delay measurements in the solar system is the Cassini time delay experiment (Bertotti *et al*., 2003). The Cassini experiment was carried out between 6 June and 7 July 2002, when the spacecraft was on its way to Saturn, around the time of a solar conjunction. The conjunction—at which the spacecraft (at a geocentric distance of 8.43 AU), the Sun and the Earth were almost aligned, in this order—occurred on 21 June 2002, with a impact parameter *b* of 1.6 solar radius, and no occultation. At this time, there is a maximum two-way general relativistic Shapiro time delay of 260 μs. The variation of this time delay, through 18 Doppler frequency measurements along spacecraft passages was verified to $0.5 \times (2.1 \pm 2.3) \times 10^{-5}$.

### 3.1.2 Ranging accuracy

The Ka band microwave ranging accuracy is a few meters. The current Lunar Laser Ranging (LLR) and Satellite Laser Ranging (SLR) accuracy with two colors is about 1 mm (Samain 1998; Murphy 2008). This 1 mm accuracy could be extended to the whole solar system (Ni, 2009).

Interplanetary laser ranging was demonstrated by MESSENGER (MErcury Surface, Space ENvironment, GEochemistry, and Ranging) (Smith *et al*., 2006). The MESSENGER spacecraft, launched on 3 August 2004, is carrying the Mercury Laser Altimeter (MLA). Between 24 May, 2005 and 31 May, 2005 in an experiment performed at about 24 million km before an Earth flyby, the MLA on board MESSENGER spacecraft linked optically with GSFC's 1.2 m telescope station. Pulses were successfully exchanged between the two terminals. From this two-way laser link, the range as a function of time at the spacecraft over $2.39 \times 10^{10}$ m (~ 0.16 AU) was determined to $\pm$ 0.2 m ($\pm$ 670 ps): a fractional accuracy of better than $10^{-11}$. A one-way uplink experiment was conducted by the same team to the Mars Orbiter Laser Altimeter on board the Mars Global Surveyor spacecraft orbiting Mars.

These interplanetary laser ranging demonstrations are encouraging. Dedicated laser ranging payload using current technology will reach the accuracy of 1 mm as in LLR/SLR. For example, the aimed accuracy of ASTROD I (Braixmaier *et al*., 2012) and Super-ASTROD (Ni, 2009) is at 1 mm.

## 3.2 Testing cosmological models

### 3.2.1 DSSY modified gravity with a quadratic Weyl term

Recently, there has been an interest in inflation and modified gravity with a quadratic Weyl term with parameter $\gamma_W$ added to the general-relativistic action (Deruelle *et al*., 2011). We have studied the solar-system tests of this theory by first deriving linearized equation of motion in the weak field limit, solving it for isolated system in the slow motion limit, and using it to derive the light propagation equations and obtaining the relativistic Shapiro time delay & light deflection(Ni, 2012). Applying these results to the solar-system measurements, we obtain constraints on the Weyl term parameter $\gamma_W$; the most stringent constraint, which comes from the Cassini relativistic time delay experiment, is for $|\gamma_W|$ to be less than $1.5 \times 10^{-3}$ AU$^2$, or $|\gamma_W|^{1/2}$ less than 0.039 AU (19 s) (Ni, 2012). Analysis of precision laboratory gravity experiments put further limit on the Weyl term parameter $\gamma_W$ to below the laboratory scale. We note in passing that there is an issue that the standard perturbation method may not find all the solutions (Deruelle 2012); non perturbative solutions should also be looked for to see how they fit the solar system experiments and give constraints on $\gamma_W$.

### 3.2.2 DGP theory and de Rham-Gabadadze massive gravity

In a five-dimensional braneworld model (DGP gravity) developed by Dvali, Gabadadze and Porrati (2000), the standard model (matter) interactions are constrained to a four-dimensional brane while gravity is modified at large distances by the arrested leakage of gravitons off our four-dimensional universe. DGP gravity has a crossover scale $r_c \approx 5$ Gpc, above which gravity becomes 5-dimensional. The model is able to produce cosmic acceleration without invoking dark energy. Lue and Starkman (2003) showed that orbits near a mass source suffer a universal anomalous precession dω/dt as large as ±5 μas/year, dependent only on the graviton's effective linewidth and the global geometry of the full, five-dimensional universe

$$|d\omega/dt| = 3c/8r_c = 5\times10^{-4} (5 \text{ Gpc}/r_c) \text{ arcsec/century}. \tag{4}$$

Iorio (2005, 2007) extended this equation to second order in eccentricity and used solar-system observations to constrain the anomalous gravitational effects. Battat, Stubbs and Chandler (2008) noticed that single point measurement uncertainties in the ranging data to Mercury and Mars are 10 m and 5-40 m, respectively, and for DGP-like precession the constraint is $|d\omega/dt| < 0.02$ arcsec/century. Therefore at the level of 0.02 arcsec/century, there is no evidence for a universal precession in excess of general relativity prediction.

One reason that the present constraints from the planetary motions are so relaxed is that they are nearly coplanar and for coplanar motion, universal precession cannot be detected using concentric relative motions. Since lunar orbit is inclined to ecliptic plane, with the progress of LLR and grand fitting together with Mars ranging, the DGP off-plane relative precession should be detectable or constrained in the future. Super-ASTROD has been proposed with one spacecraft orbit nearly vertical to the ecliptic plane with a 5 AU solar orbit and, therefore, is ideal for this measurement. Two-wavelength laser ranging through the atmosphere of Earth achieved 1 mm accuracy (Murphy, 2008). With a single point ranging accuracy of 1 mm using pulse ranging, the DGP effect of 180 m (for a mission of 10 years: $5\times10^{-5}$ arcsec $\times$ 4.8 $\times$ $10^{-6}$ rad/arcsec $\times$ 5 AU $\approx$ 180 m) for Super-ASTROD can be measured to $10^{-4}$ or better. For Super-ASTROD, 2nd order eccentricity effect in DGP theory can also be measured. This is an example of the capability of testing relativistic gravity in the future using laser ranging in the solar system (Ni, 2009).

De Rham-Gabadadze massive gravity (de Rham and Gabadadze, 2010) induces similar precessions (Gabadadze 2012). However, the crossover scale is about ten times larger, and the precession is ten times slower. Nevertheless Super-ASTROD has the capability of determining this rate to about 0.1 % or better.

This work is based on a plenary talk presented at ICGAC10 (Xth International Conference on Gravitation, Astrophysics and Cosmology), Quy-Nhon, Vietnam, and is supported in part by the National Science Council under Grants No. NSC100-2119-M-007-008 and No. NSC100-2738-M-007-004.

## References

J. M. Bardeen and J. A. Petterson (1975), *Astrophys. J*. **195**, L65-L67.
J. B. R. Battat, C. W. Stubbs and J. F. Chandler (2008), *Phys. Rev. D* **78** 022003.
B. Bertotti, *et al*. (2003), *Nature* **425,** 374-376; and ref.'s therein.
C. Braxmaier *et al.* (2012), Astrodynamical Space Test of Relativity using Optical Devices I (ASTROD I)—a class-M fundamental physics mission proposal for cosmic vision 2015–2025: 2010 Update, *Experimental Astronomy*. DOI: 10.1007/s10686-011-9281-y Online First™; arXiv:1104.0060.
I. Ciufolini and E. C. Pavlis (2004), *Nature* **431**, 958.

C. de Rham and G. Gabadadze (2010), *Phys. Rev. D* **82**, 044020.
W. de Sitter (1916), *Mon. Not. R. Astron. Soc.* **77**, 155.
N. Derulle, M. Sasaki, Y. Sendouda and A. Youssef (2011), *JCAP*, **3**, 040.
N. Deruelle, M. Sasaki and Y. Sendouda (2008), *Prog. Theor. Phys.* **119** 237-251.
N. Deruelle (2012), private communication.
G. Dvali, G. Gabadadze and M. Porrati (2000), *Phys. Lett. B*, **48**5, 208.
C. W. F. Everitt *et al*. (2011), *Phys. Rev. Lett.* **106**, 221101; arXiv:1105.3456.
E. Fomalont, *et al*. (2009), *Astrophys. J.* **699,** 1395–1402.
G. Gabadadze (2012), private communication.
F. W. Hehl and Yu. N. Obukhov (2008), *Gen. Rel. Grav.* **40**, 1239-1248.
L. Iorio (2005), *Class. Quantum Grav.* **22,** 5271.
L. Iorio (2007), *Adv. High Energy Phys.* 90731 (http://www.hindawi.com/).
L. Iorio (2009), *Space Science Reviews* **148**, 363.
V. A. Kostelecky and M. Mewes (2002), *Phys. Rev. D* **66**, 056005.
C. Lämmerzahl and F. W. Hehl (2004), *Phys. Rev. D* **70**, 105022.
S. B. Lambert and C. Le Poncin-Lafitte (2009), *Astron & Astrophys.*, **499** 331–335.
J. Lense and H. Thirring (1918), *Phys. Z.* **19**, 156.
A. Lue and G. Starkman (2003), *Phys. Rev. D* **67,** 064002.
B.-Q. Ma (2012), New Perspective on Space and Time from Lorentz Violation, arXiv:1203.5852; and ref's therein.
T. W. Murphy *et al*. (2008) *Publ. Astron. Soc. Pac.* **120,** 20-37
W.-T. Ni (1973), A Nonmetric Theory of Gravity, preprint, Montana State University [http://astrod.wikispaces.com/].
W.-T. Ni (1974), *Bull. Am. Phys. Soc.* **19**, 655.
W.-T. Ni (1977), *Phys. Rev. Lett.* **38,** 301–4.
W.-T. Ni (1984), Equivalence Principles and Precision Experiments, *Precision Measurement and Fundamental Constants II*, B. N. Taylor and W. D. Phillips, (Ed.), Natl. Bur. Stand. (U S) Spec. Publ. **617,** pp 647-651; and ref's therein.
W.-T. Ni (2005), *Int. J. Mod. Phys. D* **14**, 901-921; and ref's therein.
W.-T. Ni (2009), *Class. Quantum Grav.* **26,** 075021
W.-T. Ni (2010), *Rep. Prog. Phys.* **73**, 056901.
W.-T. Ni (2011), *Phys. Rev. Lett.* **107**, 051103.
W.-T. Ni (2012), Solar-system tests of the inflation model with a Weyl term arXiv:1203.2465.
R. D. Peccei and H. R. Quinn (1977), *Phys. Rev. Lett.* **38**, 1440-1443.
É. Samain *et al*., *Astron. Astrophys. Suppl.* **130**, 235 (1998).
L. I. Schiff (1960), *Phys. Rev. Lett.* **4**, 215.
S. S. Shapiro *et al*. (2004), *Phys. Rev. Lett.* **92,** 121101(4); and references therein.
D.E. Smith *et al*. (2006), *Science* **311**, 53.
K. S. Thorne (1988), in *Near Zero: New Frontiers of Physics*, edited by J. D. Fairbank *et al*. (W.H. Freeman and Co., New York), p. 573; and ref's therein.
S. Weinberg (1978), *Phys. Rev. Lett.* **40**, 233.
F. Wilczek (1978), *Phys. Rev. Lett.* **40**, 279.
J. G. Williams *et al*. (2004), *Phys. Rev. Lett*. **93**, 261101.